# A Framework for Cloud Security Risk Management Based on the Business Objectives of Organizations

Ahmed E. Youssef[1,2]
[1]College of Computer and Information Sciences
King Saud University, Riyadh, Saudi Arabia
[2]Faculty of Engineering, Helwan University, Cairo, Egypt
ahyoussef@ksu.edu.sa

*Abstract*— Security is considered one of the top ranked risks of Cloud Computing (CC) due to the outsourcing of sensitive data onto a third party. In addition, the complexity of the cloud model results in a large number of heterogeneous security controls that must be consistently managed. Hence, no matter how strongly the cloud model is secured, organizations continue suffering from lack of trust on CC and remain uncertain about its security risk consequences. Traditional risk management frameworks do not consider the impact of CC security risks on the business objectives of the organizations. In this paper, we propose a novel Cloud Security Risk Management Framework (CSRMF) that helps organizations adopting CC identify, analyze, evaluate, and mitigate security risks in their Cloud platforms. Unlike traditional risk management frameworks, CSRMF is driven by the business objectives of the organizations. It allows any organization adopting CC to be aware of cloud security risks and align their low-level management decisions according to high-level business objectives. In essence, it is designed to address impacts of cloud-specific security risks into business objectives in a given organization. Consequently, organizations are able to conduct a cost-value analysis regarding the adoption of CC technology and gain an adequate level of confidence in Cloud technology. On the other hand, Cloud Service Providers (CSP) are able to improve productivity and profitability by managing cloud-related risks. The proposed framework has been validated and evaluated through a use-case scenario.

## I. INTRODUCTION

The importance of Cloud Computing (CC) is increasing and it is receiving a growing interest by many scientific and business organizations [11]. According to the National Institute of Standards and Technology (NIST) [32], cloud computing is a model for enabling convenient, ubiquitous, on-demand access to a shared pool of configurable resources (e.g., networks, servers, storage, and applications) which can be easily delivered with different types of service provider interaction that follow a simple Pay-As-You-Go (PAYG) model. In PAYG model, the Cloud Service Consumers (CSC) can request the computing services as needed to their business; the services are provided on-demand by the Cloud Service Providers (CSP), and the CSC only pay for the services they have used. The many advantages that CC brings to organizations, such as high scalability and flexibility, excellent reliability and availability, economy of scale, consolidation and energy saving, are well-documented [35]. Furthermore, CC is poised to be a significant growth area, according to Forbes, CC market is projected to reach $411B by 2020 [30]. LogicMonitor has conducted a survey to explore the landscape for cloud services in 2020, one of the interesting findings in this survey is that 83% of enterprise workloads will be in the Cloud by 2020 [36].

Although the benefits of CC are significant for many organizations, it has brought many risks that influence its confidence and feasibility. Figure 1 shows the most important risks for organizations adopting CC [36]. Security is considered one of the top ranked risks of CC [12,13], From the CSC perspective, the main reasons for distrust on CC are its multi-tenancy nature and the outsourcing of sensitive data, critical applications and infrastructure onto the cloud. On the other hand, from CSP perspective, security issue in CC is also a challenge because of the complexity of the cloud model that results in a large number of heterogeneous security controls that must be consistently managed.

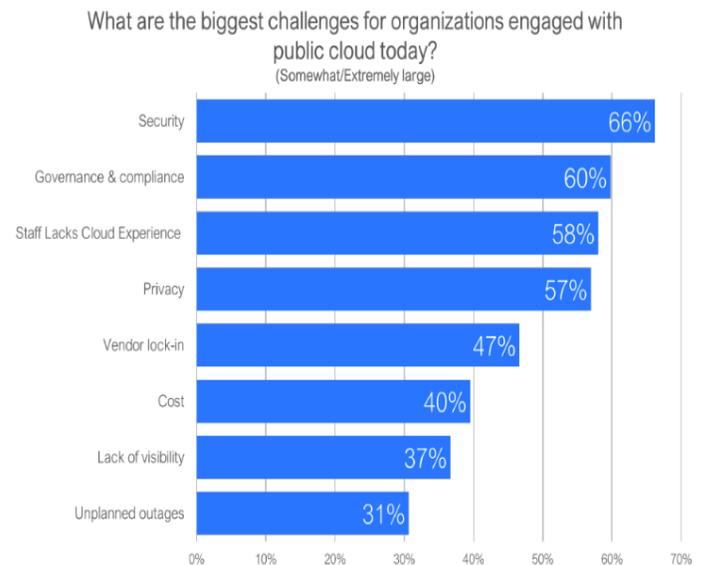

Fig.1. The biggest challenges for organization engaged with CC

Organizations have many security concerns about migration to the cloud such as loss of control over their data, lack of security guarantees, and sharing their data with malicious users. These risks often create fears in the side of organizations causing them to rethink their decisions in adopting CC technology. No matter how strongly the cloud model is secured, organizations continue suffering from lack of trust on cloud and remain uncertain about its economic feasibility. Although the provision of zero-risk service is not practically possible, an effective security risk management framework may lead to a higher confidence of organizations in CC and help them take well-informed decisions regarding the adoption of this emerging technology. Traditional risk management frameworks do not fit CC well due the assumption by those frameworks that the assets are owned and fully managed by the organization itself. Moreover, none of them considers organization's security requirements and the effect of CC security risks on its business objectives.

This paper proposes a novel Cloud Security Risk Management Framework (CSRMF) that helps organizations and CSP identify, analyze, evaluate security risks in CC platforms, and establish the best course of action to avoid or mitigate them. Unlike traditional risk management framework, CSRMF considers organization's security requirements and is driven by the impact of CC security risks on the achievement of its business objectives. It allows any organization adopting CC to be aware of cloud security risks and align their low-level management decisions according to high-level business objectives. In essence, it is designed to address impacts of cloud-specific security risks into business objectives in a given organization. Consequently, organizations are able to conduct a cost-value analysis and take a well-informed decision regarding the adoption of CC technology. On the other hand, CSP are able to improve productivity and profitability by managing cloud-related risks. This framework provides an adequate level of confidence in CC for organizations and a cost-effective productivity for CSP.

The rest of this paper is organized as follows: section 2 briefly introduces the main concepts in risk management. In section 3, related work is reviewed. Section 4 describes the proposed framework (CSRMF) in detail. In Section 5, we evaluate the framework through a use case scenario. Finally, in section 6, we give our conclusions and future work.

## II. RISK MANAGEMENT

Risk is defined as the possibility of a hazardous event occurring that will have an adverse consequence on the achievement of the objectives of an organization [2]. Risks are unavoidable and persistently exist in our daily life in almost every situation [10]. The main concepts related to risks are: *Asset:* something to which an organization assigns value and hence it needs protection. *Threat:* a potential undesired event that harms or reduces the value of an asset. *Vulnerability:* a flaw or deficiency that may be exploited by a threat to harm assets. *Risk likelihood:* the probability that a risk occurs. *Risk impact:* the degree by which a risk influences (i.e., causes loss of satisfaction of) an organization's objective(s). *Risk level:* the severity of a risk derived from its likelihood and impact. *Risk tolerance:* the amount of satisfaction or pleasure regarding the risk level. For example, a server is considered as an asset, a threat could be a backdoor virus attack, and a vulnerability is a virus scan not up to date. The likelihood that a computer is infected by this virus is medium, but its impact on data integrity is high [1, 2].

*Risk management* is the art and science of identifying, analyzing, evaluating and responding to risks throughout the service lifecycle. It enables an organization to recognize uncertain events that may result in unfortunate or damaging consequences and to set the best course of action to avoid or mitigate them [4,15]. However, in order to apply risk management effectively, it is vital to first identify the overall vision, mission and objectives of an organization. Risk management is about making decisions that contribute to the achievement of an organization's objectives such as costs with benefits and expectations in investing limited public resources. It protects and adds value to the organization and its stakeholders by:

- Enhancing safety and security in an organization.
- Protecting organization's assets and reputation.
- Optimizing operational efficiency.
- Supporting the achievement of organization's objectives by satisfying stakeholders' expectations and improve their confidence and trust.
- Improving decision making by comprehensive understanding of business activities in organizations.

A Risk Management Framework (RMF) is a set of components that provide the foundations for risk management throughout the organization. Figure 2 shows the evolution of RMF [37].

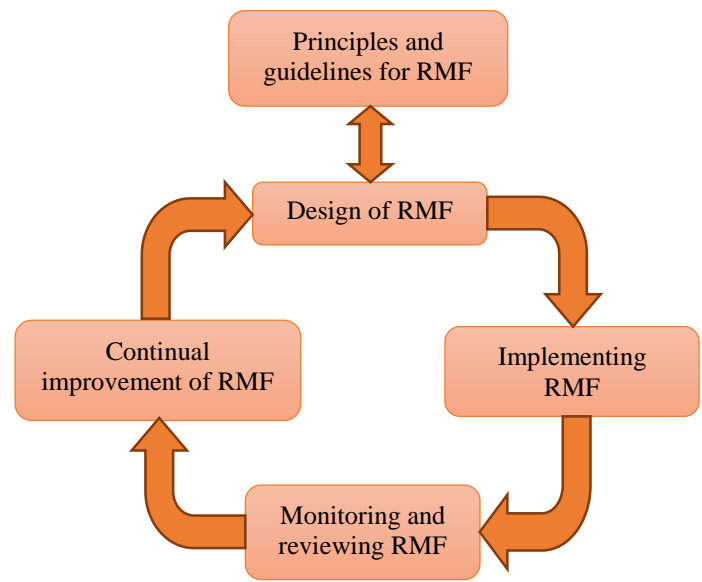

Fig. 2. Standard Risk Management Framework Evolution

## III. RELATED WORK

In literature, there are many frameworks that help in security risk management [3,5-9,14,20-29,39], however, these traditional risk management frameworks do not fit CC well due to the complexity of CC environment and the assumption by those frameworks that the assets are owned and fully managed by the organization itself. In addition, none of them considers organization's security requirements and the effect of CC security risks on the business objectives and goals of the organizations. The work presented in this paper aims to develop an RMF that is driven by the impact of CC security risks on the business objectives of organizations adopting CC technology. The existing information security risk management frameworks are described below.

QUIRC: a quantitative impact and risk assessment methodology for CC projects developed to assess the security risks associated with CC platforms [8]. This framework uses the definition of risk as a combination of the probability of a security threat event and its severity, measured as its impact. Six key security criteria (Confidentiality, Integrity, Availability, Multiparty trust, Auditability, and Usability) are identified for cloud platforms, they are referred to as the CIAMAU framework, and it is shown that most of the typical attack vectors and events map under one of these six categories. QUIRC employs a quantitative approach that gives vendors, customers and regulation agencies the ability to comparatively assess the relative robustness of different cloud vendor offerings and approaches in a defensible manner. Limitations of this approach include that it requires the meticulous collection of input data for probabilities of events, which requires collective industry inputs.

OPTIMS: an effective and efficient risk assessment framework for cloud service provision [1, 2]. Four risk categories, namely legal, technical, policy, and general were identified. This framework is beneficial for end-users and Service Providers (SP) approaching the cloud to deploy and run services, as well as Infrastructure Providers (IP) to deploy and operate those services. These benefits include supporting various parties for making informed decisions regarding contractual agreements. The risk assessment framework is fully integrated in the OPTIMIS toolkit, which simplifies cloud self-management, optimizes the cloud service lifecycle, and supports various cloud architectures. However, the SP dynamic risk assessment is limited due to the lack of support for service consumer's side monitoring tools and the limited availability of shared monitored data from IPs.

CARAM: is a qualitative and relative risk assessment model that helps CSC select CSP that fit their risk profile the best [9]. It consists of tools that complements the various recommendations of ENISA [33] and CSA [40]. These tools include a questionnaire for CSC, an algorithm to classify the answers to Cloud Assessment Initiative Questionnaire (CAIQ) to discrete values, a model that maps the answers to both questionnaires to risk values, and a multi criteria decision approach allowing to quickly and reliably compares multiple CSP. However, there are limitations that may affect the accuracy of the results mainly stemming from the analyzed input data such as: Vague formulation of the CAIQ answers provided by the analyzed CSP, Possibility for deliberate misinformation in the CAIQ, and Ineffective implementation of the security controls by the analyzed CSPs.

CRAMM: a risk analysis and management method that includes a comprehensive range of risk assessment tools that are fully compliant with ISO27001 and address tasks such as: asset dependency modeling, identifying and assessing threats and vulnerabilities, assessing risk levels, and identifying required controls [14,22]. It provides a staged and disciplined approach embracing both technical (e.g. hardware and software) and non-technical (e.g. physical and human) aspects of security. The major flaws in CRAMM are: 1) quantitative risk assessment cannot be provided. Hence, there is need to extend this methodology in this direction and 2) it does not clearly talk about the security attributes e.g. Confidentiality, Integrity, and Availability [23].

COBRA: a risk assessment model that consists of a range of risk analysis, consultative and security review tools which were developed largely in recognition of changing nature of IT and security, and the demands placed by business upon these areas [39]. The default risk assessment process usually consists of three stages: questionnaire building, risk surveying, and report generation. The major weaknesses of COBRA are 1) risk assessment technique is not clearly mentioned; hence, there is need to extend this methodology in this direction and 2) threats and vulnerabilities play a very important role in the process of risk assessment; but how these are taken into consideration, is not clearly given in COBRA [23].

## IV. THE PROPOSED FRAMEWORK

We propose a Cloud Security Risk Management Framework (CSRMF) that implies methods for identifying, analyzing, evaluating, treating, and monitoring security risks throughout the cloud service lifecycle. In this context, assets include data hosted on the cloud, physical nodes, virtual machines, and other cloud resources as well as the Service Level Agreement (SLA), risks are potential security threats attacking the assets in CC platforms causing loss of satisfaction of organization's objectives. The proposed CSRMF aims at:

- Identifying the risks that present threats to the cloud within the context of organization's concerns.
- Analyzing and evaluating identified risks with respect to organization's goals and objectives.
- Applying the best course of treatment actions to reduce the likelihood and/or the impacts of these risks.
- Monitoring the currency of identified risks regularly to ensure that treatment actions are valid.
- Establishing a dynamic relationship between the organization and CSP during risk management process to ensure the compliance to SLA.

Figure 3 shows the main components of the proposed framework. In the following subsections, we discuss each one of these components.

### A. Identifying Organization's Business Objectives

Organizational objectives are short-term and medium-term goals that an organization seeks to accomplish. Achievement of these objectives helps an organization reach its overall strategic goals. Therefore, the proposed framework, CSRMF, is driven by the organization's high-level objectives. Organizational objectives are established through understanding the overall internal culture (e.g. vision, mission, etc.) of the organization and a number of environmental analyses that include identifying the constraints and opportunities of the operating environment.

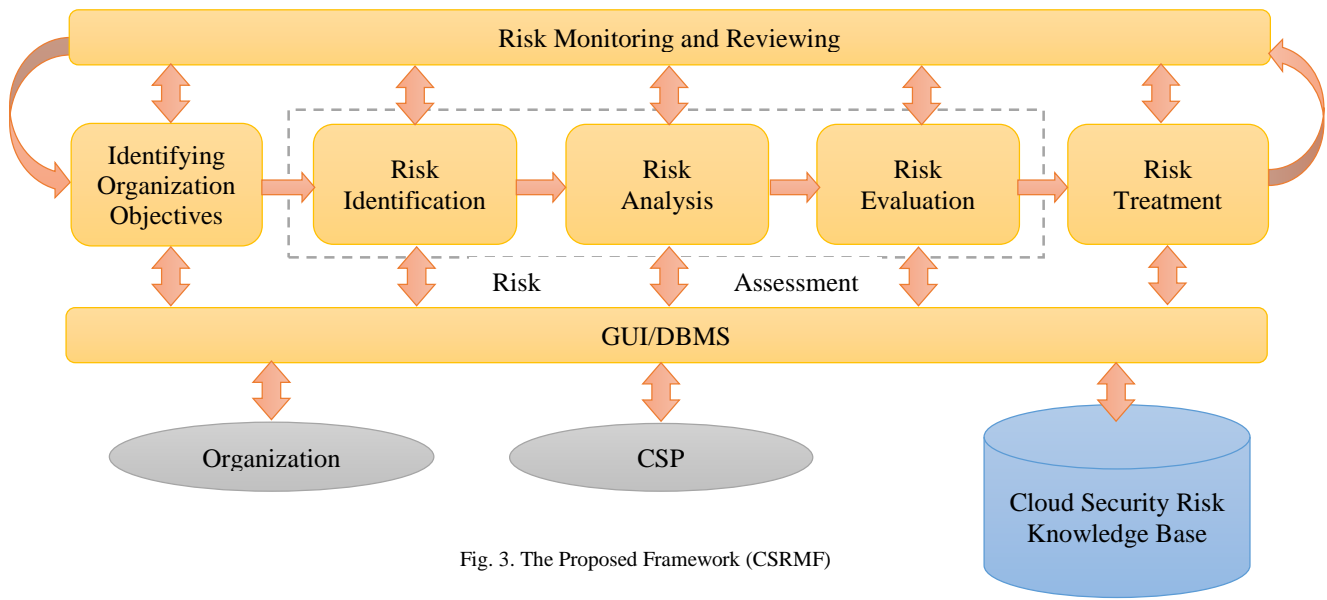

Fig. 3. The Proposed Framework (CSRMF)

To set the organization's objectives, CSRMF proposes to conduct SWOT analysis [41] where organizations identify their internal Strengths and Weaknesses as well as external Opportunities and Threats. This information allows CEO to develop objectives and strategies that are relevant and realistic to their organizations. In CSRMF, organizational objectives should follow the SMART model (i.e., should be Specific, Measurable, Attainable, Relevant, and Time bound). To apply the SMART model, CEO have to ask themselves the following questions when setting their organizations' objectives:

- **Specific** – What type of company do you want to be the best at? On what scale do you want to compete? Do you want to be the best company in your area or in the world?
- **Measurable** – How will you know when you have achieved your objective? What benchmarks are you going to use to measure your success?
- **Attainable** – Is this objective achievable given your resources? What are the obstacles that you are going to encounter and can you get past the hurdles?
- **Relevant** – How relevant is this objective to the company and its employees? Will it benefit your organization?
- **Time bound** – When do you want to achieve this objective by?

Examples for good organization objectives are: achieving financial success, increasing sales figures, improving human resources, retaining talented employees, focusing on customer service, and establishing brand awareness.

### B. Risk Identification

The second phase in CSRMF is to identify risks that are likely to affect the achievement of the objectives of the organization. The identification of security risks affecting cloud services in organizations that adopt CC is the most critical step in risk management. The better identifying and understanding these risks, the more meaningful and effective will be the risk management process. The appropriate risk identification method will depend on the application area (i.e., nature of activities and the hazard groups), the nature of projects in organization, resources available, regularity requirements and client requirements as to objectives, desired outcome and the required level of detail. However, there is no single scientific method that guarantees identification of all risks [10].

Risks are caused by security threats that may exploit vulnerabilities in CC platform to harm organization's assets and consequently affect the achievement of its objectives. Therefore, in order to identify risks precisely, we need to identify assets, vulnerabilities, and threats in CC platform. Since there is no single scientific approach that guarantees identification of all risks, CSRMF employs a hybrid approach that combines two techniques for risk identification. This combination will be more effective for full and adequate coverage of risks. Risk identification techniques that are employed by CSRMF are: *documented knowledge acquisition* and *brainstorming*.

*B.1. Documented Knowledge Acquisition*

This technique implies collecting and reading documents about CC risk domain such as books, surveys, articles, and regulations. Many documents in literature have attempted identifying CC risks and threats [31-33, 38,40]. One of the most useful documents regarding CC risk is the one provided by the European Network and Information Security Agency (ENISA) [33] that affords generic lists of risks for CC. Examples of such risks are Lock-in, Resource Exhaustion, Isolation Failure and Malicious Insider, a sample of these risks is shown in figure 4. However, these lists do not reflect the organization objectives nor they reveal a specific class of business applications.

R.1 LOCK-IN

| Probability | HIGH | Comparative: Higher |
|---|---|---|
| Impact | MEDIUM | Comparative: Equal |
| Vulnerabilities | V13. Lack of standard technologies and solutions | |
| | V46. Poor provider selection | |
| | V47. Lack of supplier redundancy | |
| | V31. Lack of completeness and transparency in terms of use | |
| Affected assets | A1. Company reputation | |
| | A5. Personal sensitive data | |
| | A6. Personal data | |
| | A7. Personal data - critical | |
| | A9. Service delivery – real time services | |
| | A10. Service delivery | |
| Risk | HIGH | |

Fig.4. A sample of ENISA CC Risk Identification (LOCK-IN risk)

The documented knowledge acquisition technique is an important prerequisite to other techniques. However, the huge amount of available documentation may lead to irrelevant details and outdated information. An effective solution that we employed to solve this issue is to use meta-knowledge (know what you need to know and what you do not need to know) to prune the document space. The knowledge acquired in this step is stored in a cloud security risk knowledge base for use in the next step (i.e., brainstorming).

*B.2. Brainstorming*

Brainstorming a semi-structured creative group-based activity, used most often in ad-hoc business meetings to come up with new ideas for solving problems, innovation or improvement [34]. It usually involves a group, under the direction of a facilitator and implies two stages:

1. Idea generation: generate as many ideas as possible to address the problem from each participant without criticism.
2. Idea evaluation: by all participants together according to agreed criteria (e.g. value, cost, feasibility) to prioritize ideas.

In CSRMF, members of a team that comprises information security experts and a diverse group of stakeholders in the organization meet to identify organization's assets, vulnerabilities, and potential threats. Risks identification takes place in a series of group workshops; group sessions provide a wider exploration of issues and more creative ways for identifying risks. The group uses the knowledge acquired in the previous step to identify different risks. The outcome of this step is a list of identified risks which is reviewed by an independent stakeholders group. If satisfaction is achieved the risk management process proceeds to the next phase, otherwise, it goes through another round of risk identification.

*C. Risk Analysis*

Risk analysis involves the estimation of risks likelihoods and impacts. CSRMF deploys a quantitative approach for risk analysis and assumes the following:

- Objective weight ($w_j$): the importance of an objective $o_j$, ($0 \leq w_j \leq 1$, $\sum_j w_j = 1$, $j = 1,2, \ldots, m$)
- Risk Likelihood $L(r_i)$: the probability of occurrence of risk $r_i$. ($0 \leq L(r_i) \leq 1$, $i = 1,2, \ldots, n$)
- Risk Impact $I(r_i, o_j)$: the effect of $r_i$ on $o_j$, ($0 \leq I(r_i, o_j) \leq 1$), where 0 means no loss of satisfaction in $o_j$, 1 means total loss of satisfaction in $o_j$, m and n are numbers of objectives and risks respectively.

The goal of risk analysis phase is to estimate values for $L(r_i)$ and $I(r_i, o_j)$. A widely accepted consensus-based estimation technique is the Delphi method [8, 16-19]. Three essential characteristics of Delphi method are: 1) structured and iterative information flow, 2) anonymity of the participants in order to alleviate peer pressure and other performance anxieties, and 3) iterative feedback of the participants until consensus is reached. We adapted the Delphi technique for the estimation of security risk likelihood and impacts; this is shown in figure 5.

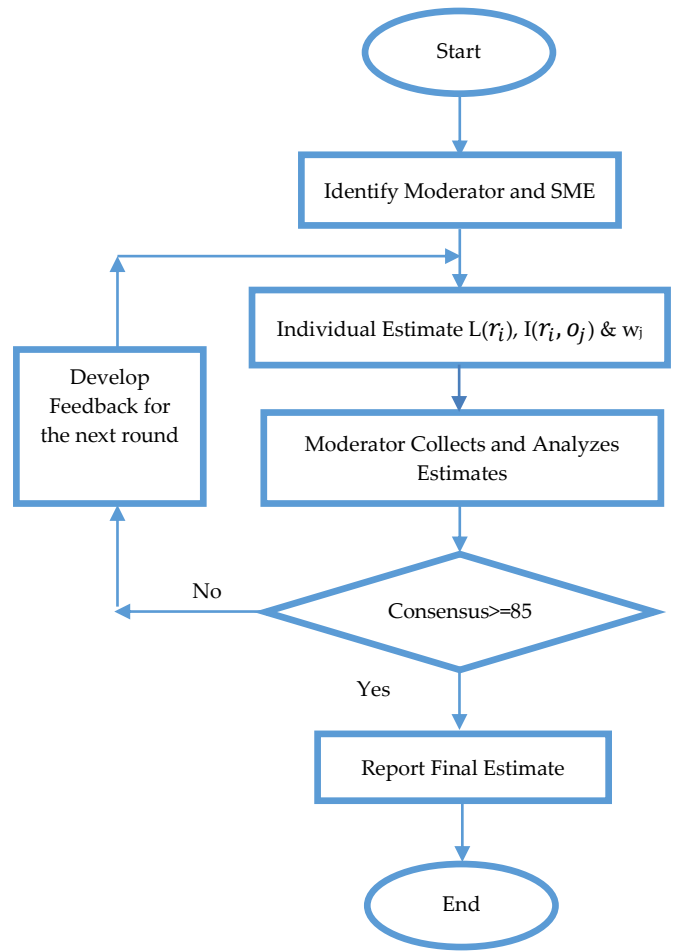

Fig. 5. The Delphi process for risk analysis as used in CSRMF

In CSRMF Delphi technique, a moderator is used to control and facilitate information gathering from a selected group of Subject Matter Experts (SME). SME are knowledgeable about the likelihoods and impacts of risks on the organization's particular type of business. During the Delphi process, each participant is asked to provide his best numerical estimates of $w_j$, $L(r_i)$ and $I(r_i, o_j)$. Following this step, the moderator collects the estimates from all participants in anonymous presentation, shares and analyses the combined results with all participants. The participant are encouraged to iteratively reconsider and modify their estimates based on the feedback from previous discussion. When estimates reach a consensus (e.g. 85% or more), the moderator reports the final estimates to be used in the next phase.

*D. Risk Evaluation*

Risk evaluation implies estimate of the risk level (i.e., risk severity) to be able to decide whether the risk is tolerable (i.e., acceptable) by the organization or not. Tolerable risk criteria must be defined, approved, and documented by relevant committee from experts and stakeholders. Should the estimated risk level greater than the tolerable level then the specific risk needs treatment or improved countermeasures. In CSRMF, risks are evaluated using a quantitative approach, the level of risk $r_i$ (i.e., $Level(r_i)$) is estimated using equation 1.

$$Level(r_i) = L(r_i) \sum_{j=1}^{m} w_j\, I(r_i, o_j) \quad (1)$$

$$0 \leq Level(r_i) \leq 1$$

Risk level ($Level(r_i)$) ranges between 0 and 1, where 0 means $r_i$ has no effect (min. severity) on the organization's objectives and 1 means $r_i$ has significant effect (max. severity) on organization's objectives. A risk, $r_i$, may be considered acceptable (tolerable) if $Level(r_i)$ is less than threshold $\alpha$, otherwise $r_i$ requires treatment. This threshold ($0 \leq \alpha \leq 1$) is predetermined by the organization. By applying this condition an organization can achieve an acceptable Global Risk Level (GRL) which is given by equation 2.

$$GRL = \sum_{i=1}^{n} Level(r_i) \quad (2)$$

*E. Risk Treatment*

Any unacceptable risk should be treated, which means to reduce its risk level to become less than the threshold $\alpha$. The objective of risk treatment is to develop cost effective options for treating unacceptable risks. Different treatment options may be employed, which are not necessarily mutually exclusive or appropriate in all circumstances such as risk avoidance, risk transfer or share with a third party and risk mitigation (reduction) which means controlling the likelihood of risk occurrence, or controlling the impact of the consequences if the risk occurs.

CSRMF employs risk mitigation approach for risk treatment. The ultimate goal is to reduce GRL by reducing $Level(r_i)$ for each unacceptable risk. Risk likelihood, $L(r_i)$, can be reduced through preventative maintenance, or quality assurance and management, change in business systems and processes. On the other hand, risk impact, $I(r_i, o_j)$, can be reduced through contingency planning, minimizing exposure to sources of risk or separation/relocation of an activity and resources. Risk mitigation actions can be determined using a combination of documented knowledge acquisition and brainstorming techniques. Examples for CC risk mitigation actions (countermeasures) adopted in CSRMF are shown in Table 1.

TABLE 1: EXAMPLES OF RISK COUNTERMEASURES USED IN CSRMF

| CC risk | Countermeasures |
|---|---|
| Account or service hijacking | -Identify and access management guidance<br>-Dynamic credentials |
| Data leakage | -Fragmentation Redundancy Scattering (FRS)<br>-Digital signature<br>-Encryption |
| Customer data manipulation | -Web application scanners |
| Malicious VM | -Protecting aegis from live migrations of VMs |
| Sniffing/spoofing virtual Net | -Virtual Network security guarantees |

*F. Risk Monitoring*

The last phase in CSRMF is to monitor and evaluate the effectiveness of the preferred risk treatments and current control activities. To do this, we need to estimate the risk level reduction after applying a countermeasure technique. Suppose that $c_k$ (k=1,2,3, …) is a countermeasure that can be applied to mitigate a risk (i.e., reduce its level). The Delphi approach described in section 4.C can be used to estimate risk level reduction of $r_i$ after applying $c_k$ which is denoted as $LevelRed(r_i|c_k)$. $LevelRed(r_i|c_k)$ is a measure to the amount by which a countermeasure $c_k$ mitigates (reduces the level of) risk $r_i$. Its value ranges between 0 and 1 ($0 \leq LevelRed(r_i|c_k) \leq 1$) where 0 means no reduction, 1 means risk elimination.

The *Combined Risk Reduction* (CRR) of a risk $r_i$ which measures the resultant (i.e., joint) mitigation in $r_i$ after applying a course of countermeasures is given by equation 3. Its value ranges between 0 and 1 ($0 \leq CRR(r_i) \leq 1$) where 0 means no reduction, 1 means risk elimination. This metric is used to decide whether a treatment course for a risk is successful or not.

$$CRR(r_i) = 1 - \prod_{k=1}^{p}(1 - LevelRed(r_i|c_k)) \quad (3)$$

$$0 \leq CRR(r_i) \leq 1$$

(p is the number of countermeasures applied to $r_i$)

The Global Risk Reduction (GRR) for the organization is given by equations (6).

$$GRR = \sum_{i=1}^{n} CRR(r_i) \quad (4)$$

V. FRAMEWORK VALIDATION AND EVALUATION

In order to validate the proposed framework for usability and applicability, we provide a step-by-step use-case scenario that shows how an organization can benefit from the proposed framework to manage Cloud security risks. Advanced Telecom (AT) is a leading telecommunications company that has a broad range of customers. AT's services bundle includes fixed landlines, Internet and mobile communications. AT employs 80 thousand employees, who spare no effort or time to reach customers and provide best services. The CEO of AT thought that it would be a great idea to develop several Intranet site applications that would allow employees in AT to share their knowledge. He also thought it would make sense to make some information available to the company's clients. For example, the company could provide advertisements about products, articles, links to other sites, and an Ask the Expert feature to help build relationships with current and future clients. He has heard about the cutting-edge CC technology and thought that it would probably be a good idea to adopt the Cloud technology in his company to support the Intranet project; however, he was worried about the security risks associated with that technology. Since AT emphasizes the importance of high-payoff projects, he wanted to explore the management of security risks in CC environment before adopting this technology in his company. Our goal is to help AT company take a decision on the adoption of CC using our proposed CSRMF.

*A. Phase 1: Identifying Organization's Objectives*

AT uses SWOT analysis and SMART model to help identify its business objectives. AT's representatives would provide the underlying information on AT's business objectives and the security requirements to protect these objectives against security risks as well as information concerning risk tolerance criteria. This Information is stored in the risk knowledge base and is used as a profile for the organization. The outcomes of this phase are shown in tables 2 and 3.

TABLE 2: BUSINESS OBJECTIVES FOR AT ORGANIZATION

| Symbol | Objective ($o_j$) |
|--------|-------------------|
| $o_1$  | Enhance customer trust and build relationships with current and future customers |
| $o_2$  | Boost employees' relationships and allow knowledge share among them |
| $o_3$  | Provide perfect customer services and improve customer satisfaction |
| $o_4$  | Increase profitability and decrease operational costs |

TABLE 3: SECURITY REQUIREMENTS FOR AT ORGANIZATION

| Security requirements | Confidentiality – medium<br>Integrity – high<br>Availability – high |
|-----------------------|---------------------------------------------------------------------|
| Risk tolerance        | 0.25                                                                |

*B. Phase 2: Risk Identification*

A team consists of seven members of information security experts (i.e., SME) and a diverse group of stakeholders in AT uses documented knowledge such as those described in section 4.B to gather information on security risks related to CC that are likely to affect the organization's objectives. Information regarding identified risks are stored in the risk knowledge base. The team then meets, conducts brainstorming session, and uses the knowledge stored in the risk knowledge base to prepare a final list of possible risks. This list is shown in table 4.

TABLE 4: IDENTIFIED RISKS FOR AT ORGANIZATION

| Symbol | Risks |
|--------|-------|
| $r_1$  | Account hijacking |
| $r_2$  | Data leakage |
| $r_3$  | Denial of services |
| $r_4$  | Insecure VM migration |
| $r_5$  | Sniffing/spoofing virtual networks |

*C. Phase 3: Risk Analysis*

The team utilizes the Delphi technique explained in section 4.C to estimate values for the weights, $w_j$, risk likelihoods, $L(r_i)$, and risk impacts, $I(r_i, o_j)$. These information are shown in table 5. For example, the weight of $o_2$ is $w_2 = 0.2$, the likelihood of $r_3$ is $L(r_3) = 0.5$, while the impact of $r_3$ on $o_2$ is $I(r_3, o_2) = 0.3$ (all shaded in gray).

TABLE 5: RISK IMPACT MATRIX FOR AT ORGANIZATION

| ↓ $w_j$   $L(r_i)$ → | $r_1$ /0.6 | $r_2$ /0.2 | $r_3$ /0.5 | $r_4$ /0.7 | $r_5$ /0.3 |
|---------------------|------------|------------|------------|------------|------------|
| $o_1$ / 0.2         | 0.65       | 0.15       | 0.4        | 0.85       | 0.1        |
| $o_2$ / 0.2         | 0.85       | 0.35       | 0.3        | 0.8        | 0.3        |
| $o_3$ / 0.3         | 0.75       | 0.8        | 0.25       | 0.7        | 0.7        |
| $o_4$ / 0.3         | 0.8        | 0.65       | 0.1        | 0.6        | 0.2        |

*D. Phase 4: Risk Evaluation*

The levels of identified risks are evaluated using equations 1, the results are shown in table 6. This evaluation allows the organization to decide whether the risk is tolerable (i.e., acceptable) or not. Tolerable risk criteria have been defined, approved, and documented by the relevant committee of experts and stakeholders in phase 1. The committee has agreed that the risk level for a tolerable risk should not exceed 0.25 (i.e., $\alpha = 0.25$), this means that $r_1$ and $r_4$ need treatment to lower their risk levels below 0.25. The GRL (the sum of all risk levels) has been estimated using equation 2.

TABLE 6: RISK LEVELS FOR AT ORGANIZATION

| $r_i$ | $Level(r_i)$ |
|-------|--------------|
| $r_1$ | **0.46**     |
| $r_2$ | 0.11         |
| $r_3$ | 0.12         |
| $r_4$ | **0.50**     |
| $r_5$ | 0.11         |
| GRL   | 1.30         |

*E. Phase 5: Risk Treatment*

Unacceptable risks ($r_1$, $r_4$) require treatment; the objective of this phase is to identify countermeasures to mitigate unacceptable risks. The ultimate goal is to reduce GRL for the organization. Risk countermeasures are identified by the team using a combination of knowledge acquisition and brainstorming techniques. Countermeasures used by AT for $r_1$ and $r_4$ are listed in table 7.

TABLE 7: RISK COUNTERMEASURES EMPLOYED BY AT ORGANIZATION

| Symbol | Countermeasure used to mitigate risks | Risk Mitigated |
|---|---|---|
| $c_1$ | Identify and access management guidance | $r_1$ |
| $c_2$ | Dynamic credentials | $r_1$ |
| $c_3$ | Protecting aegis from live migrations of VMs | $r_4$ |

*F. Phase 6: Risk Monitoring*

Using the Delphi technique, the team would estimate $LevelRed(r_i|c_k)$ for $r_1$ and $r_4$ as per table 7. This is given in the risk reduction matrix shown below in table 8. For each unacceptable risk, the risk reduction matrix shows risk reduction by each alternative countermeasure and estimates its CRR as per equation 3.

TABLE 8: RISK REDUCTION MATRIX FOR AT ORGANIZATION

| $c_k$ | $LevelRed(r_1|c_k)$ | $LevelRed(r_4|c_k)$ |
|---|---|---|
| $c_1$ | 0.8 | 0 |
| $c_2$ | 0.9 | 0 |
| $c_3$ | 0 | 0.9 |
| $CRR(r_i)$ | **0.98** | **0.9** |

From table 8, we can see that $CRR(r_1) = 0.98$ which means that the new risk level of $r_1$ after treatment is $0.46*(1-0.98) = 0.01 < 0.25$ and $CRR(r_4) = 0.9$ which means that the new risk level of $r_4$ after treatment is $0.50*(1-0.9) = 0.05 < 0.25$. The new value of GRL after treatment =0.40, compared to 1.30 before treatment with %69 risk reduction, this is shown in table 9. The global risk reduction in AT organization GRR= 0.98+0.9= 1.88. Finally, the organization should continuously monitor the occurrence of the identified risks to ensure that the treatment actions are still valid and to identify new risks that may occur.

TABLE 9: RISK LEVELS FOR AT ORGANIZATION

| $r_i$ | Risk Level | |
|---|---|---|
| | Before mitigation | After mitigation |
| $r_1$ | **0.46** | 0.01 |
| $r_2$ | 0.11 | 0.11 |
| $r_3$ | 0.12 | 0.12 |
| $r_4$ | **0.50** | 0.05 |
| $r_5$ | 0.11 | 0.11 |
| GRL | **1.30** | **0.40** |

VI. CONCLUSION AND FUTURE WORK

CC offers numerous advantages to organizations in terms of economical saving, elasticity, flexibility, and minimal management effort. However, security and privacy concerns of CC have always been the focus of the impediments to its widespread adoption by businesses. Over time, organizations tend to relax security risks associated with CC, however, this relaxation requires a regular effective security risk management. In this paper, we proposed a novel framework for cloud security risk management that helps organizations and CSP identify, analyze, evaluate, and mitigate security risks in their CC platforms. It allows any organization adopting CC to be aware of cloud security risks and align their low-level management decisions according to high-level business objectives. In essence, it is designed to address impacts of cloud-specific security risks into business objectives in a given organization. Consequently, organizations are able to conduct a cost-value analysis and take a well-informed decision regarding the adoption of CC technology. On the other hand, CSP are able to improve productivity and profitability by managing cloud-related risks. This framework provides an adequate level of confidence in CC for organizations and a reliable and cost-effective productivity for CSP. In the future, we plan to explore quantitative techniques based on statistical analysis for risk management in CC so that we can reach a higher level of confidence in this emerging technology for organizations.